\documentclass{sig-alternate}
\newfont{\mycrnotice}{ptmr8t at 7pt}
\newfont{\myconfname}{ptmri8t at 7pt}
\let\crnotice\mycrnotice%
\let\confname\myconfname%

\usepackage[german,american]{babel}
\usepackage[T1]{fontenc}
\usepackage[utf8]{inputenc}
\usepackage{times}
\usepackage{textcomp}
\usepackage{graphicx}
\usepackage{amssymb}
\usepackage{siunitx}
\usepackage{microtype} 

\usepackage{url}
\renewcommand{\tt}{\fontencoding{OT1}\fontfamily{cmtt}\selectfont}

\usepackage[usenames,dvipsnames]{color}
\usepackage{booktabs}
\usepackage[numbers,sectionbib]{natbib}
\makeatletter
\def\@copyrightspace{\relax}
\makeatother

\begin{document}

\title{SWDE : A Sub-Word And Document Embedding Based Engine for Clickbait Detection\titlenote{The first five authors have equal contribution.}}
\numberofauthors{7}
\author{
\alignauthor
Vaibhav Kumar\\
\affaddr{International Institute of Information Technology Hyderabad}\\
\affaddr{vaibhav.kumar@research.iiit.ac.in}\\
\alignauthor
Dhruv Khattar\\
\affaddr{International Institute of Information Technology Hyderabad}\\
\affaddr{dhruv.khattar@research.iiit.ac.in}\\
\alignauthor
Mrinal Dhar\\
\affaddr{International Institute of Information Technology Hyderabad}\\
\affaddr{mrinal.dhar@research.iiit.ac.in}\\
\alignauthor
\and
Yash Kumar Lal\titlenote{The author was an intern at International Institute of Information Technology Hyderabad when this work was done.}\\
\affaddr{Manipal Institute of Technology, Manipal}\\
\affaddr{yash.kumar4@learner.manipal.edu}\\
\alignauthor
Abhimanshu Mishra\\
\affaddr{Birla Institute of Technology, Mesra}\\
\affaddr{abhimanshu.mishra@gmail.com}\\
\alignauthor
Vasudeva Varma\\
\affaddr{International Institute of Information Technology Hyderabad}\\
\affaddr{vv@iiit.ac.in}\\
\alignauthor
\and
Manish Shrivastava\\
\affaddr{International Institute of Information Technology Hyderabad}\\
\affaddr{m.shrivastava@iiit.ac.in}\\
}

\maketitle

\begin{abstract}
In order to expand their reach and increase website ad revenue, media outlets have started using clickbait techniques to lure readers to click on articles on their digital platform. Having successfully enticed the user to open the article, the article fails to satiate his curiosity serving only to boost click-through rates. Initial methods for this task were dependent on feature engineering, which varies with each dataset. Industry systems have relied on an exhaustive set of rules to get the job done. Neural networks have barely been explored to perform this task. We propose a novel approach considering different textual embeddings of a news headline and the related article. We generate sub-word level embeddings of the title using Convolutional Neural Networks and use them to train a bidirectional LSTM architecture. An attention layer allows for calculation of significance of each term towards the nature of the post. We also generate Doc2Vec embeddings of the title and article text and model how they interact, following which it is concatenated with the output of the previous component. Finally, this representation is passed through a neural network to obtain a score for the headline. We test our model over 2538 posts (having trained it on 17000 records) and achieve an accuracy of 83.49\% outscoring previous state-of-the-art approaches.
\end{abstract}

\section{Introduction}

In recent years, content delivery has changed drastically, shifting from offline methods to the Internet. It is now the primary source of information for a majority of the populace, especially for ever-changing news updates. This has also caused a shift in users' preferred sources. Previously, these preferences were static, sticking to a particular news source. Now, with the plethora of information available easily, there is no differentiation in the source it has been gathered from, with users opting to go for whatever is convenient.

Keeping up with the times, news agencies have expanded their digital presence, increasing their reach exponentially. They generate revenue by (1) advertisements on their websites, or (2) a subscription based model for articles that might interest users. Since multiple agencies offer similar content, the user has his pick. To lure in more readers and increase the number of clicks on their content, subsequently enhancing their agency's revenue, writers have begun adopting a new technique - clickbait.

Merriam-Webster defines clickbait as something (such as a headline) to encourage readers to click on hyperlinks based on snippets of information accompanying it, especially when those links lead to content of dubious value or interest. It is built to create, and consequently capitalise, on the Loewenstein information gap \cite{Loewenstein} by purposefully misrepresenting or promising what can be expected while reading a story on the web, be it through a headline, image or related text.

We propose a two-pronged approach to detect such headlines. The first component leverages distributional semantics of the title text and models its temporal and sequential properties. The article title is represented as a concatenation of its sub-word level embeddings. The sub-word representation serves as input to a bidirectional LSTM network. The contribution of a sub-word towards the clickbait nature of the headline is calculated in a differential manner since the output of the LSTM is passed into an attention layer \cite{bahdanau2014neural}, following which it goes through a dense layer. The second component focuses on Doc2Vec embeddings of the title and article content, performing an element wise multiplication of the two. This is concatenated with the dense layer output from the previous component. The obtained output is then passed through multiple hidden layers which performs the final classification.

Previous work in this field that has exploited the power of embeddings has considered either word vectors, for their ability to create context-sensitive word representations, or character-level word embeddings to model the orthographic features of a word. We propose the use of sub-word level representations since it incorporates the word's morphological features. Attaching an attention mechanism to it helps us identify the surprise associated with each representation within the clickbait. One of the identifying characteristics of clickbait is that the article title differs from the text attached to it. For this reason, we define a component to capture the interaction between these attributes and augment our model.

\section{Related Work}

The importance of detecting clickbait headlines has increased exponentially in recent years. Initial work in this domain can be traced back to \cite{Biyani:2016:ASG:3015812.3015827}, relying on heavy feature engineering on a specific news dataset. These works define the various types of clickbait and focus on the presence of linguistic peculiarities in the headline text, including various informality metrics and the use of forward references. Applying such techniques over a social media stream was first attempted by \cite{potthast:2016} as the authors crowdsourced a dataset of tweets \cite{potthast:2017b} and performed feature engineering to accomplish the task. \cite{Chakraborty2016StopCD} have tried to expand the work done for news headlines they collected from trusted sources.

\cite{ankesh2017weused} used the same collection of headlines as \cite{Chakraborty2016StopCD} and proposed the first neural network based approach in the field. They employed various recurrent neural network architectures to model sequential data and its dependencies, taking as its inputs a concatenation of the word and character-level embeddings of the headline. Their experiments yielded that bidirectional LSTMs \cite{schuster1997bidirectional} were best suited for the same. \cite{PhilippeThomasCINN} built BiLSTMs to model each textual attribute of the post (post-text, target-title, target-paragraphs, target-description, target-keywords, post-time) available in the corpus \cite{potthast:2017b}, concatenating their outputs and feeding it to a fully connected layer to classify the post. Attention mechanisms \cite{bahdanau2014neural} have grown popular for various text classification tasks, like aspect based sentiment analysis. Utilising this technique, \cite{YiweiZhouCDTUSN} deployed a self-attentive bidirectional GRU to infer the importance of each tweet token and model the annotation distribution of headlines in the corpus.

Word vectors and character vectors have been used across various approaches proposed to solve this problem. However, we suggest the use of subword representations to better analyse the morphology of possible clickbait-y words. We also attempt to model the interaction between the title of an article and its text.

\begin{figure}[h]
    \centering
    \includegraphics[height=4cm]{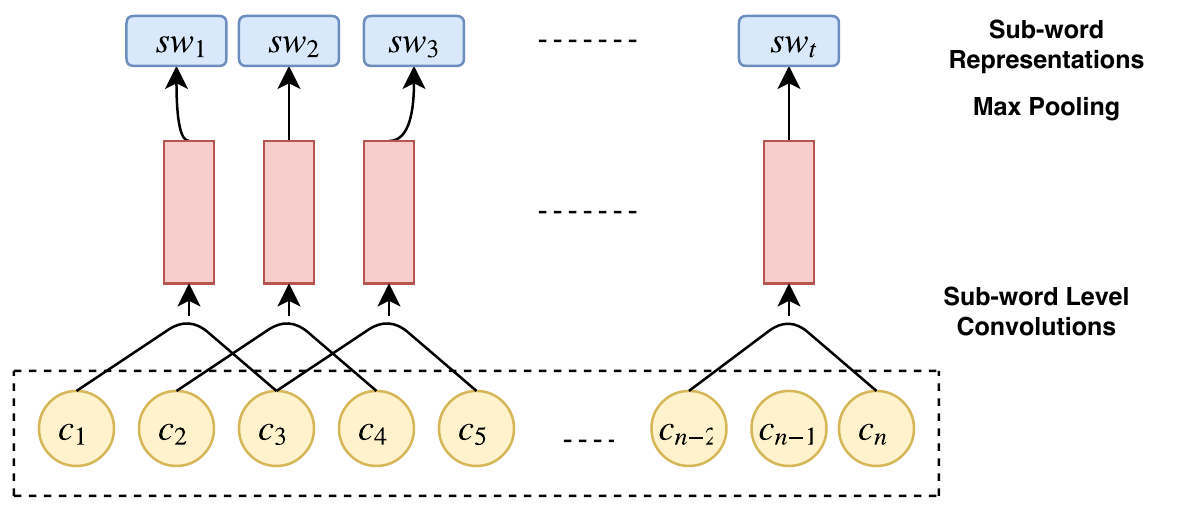}
    \caption{Architecture for learning Sub-word Level Representations using CNN}
    \label{fig:SWCNN}
\end{figure}

\section{Model Architecture}
We now describe our approach to clickbait detection and the reasons behind devising such a model. Our approach is a fusion of multiple components, each exploiting a particular type of embedding: (1) BiLSTM with attention, and (2) Doc2Vec enrichment. Figure \ref{fig:full_archi} lays out our proposed architecture.

We start with an explanation of the various types of embeddings we have used and proceed to describe the various components of our model, both individually and together. Finally, we cover how the parameters are learned.

\subsection{Sub-word Level Representation}

Word2Vec \cite{MikolovEEWRVS} has fast become the most popular text embedding method for text since it models a word based on its context. \cite{Kim:2016:CNL:3016100.3016285} proposed a convolutional neural network architecture to generate subword-level representations of words in order to capture word orthography. Sub-word level embeddings learn representations for character n-grams and represent words as the sum of the n-gram vectors \cite{DBLP:journals/corr/BojanowskiGJM16}. Such representations also take into account word roots and inflections, rather than just word context. They work well even with highly noisy text with containing misspellings due to the model learning morpheme-level feature maps. They have proven to be extremely useful in tasks such as sentiment analysis \cite{DBLP:journals/corr/PrabhuJSV16}, PoS tagging \cite{P16-2067} and language modeling \cite{Kim:2016:CNL:3016100.3016285}. These intermediate sub-word feature representations are learned by the filters during the convolution operation. We generate such an embedding by passing the characters of a sentence individually into 3 layer 1D convolutional neural network. Each filter then acts as a learned sub-word level feature. A representation for this architecture can be found in Figure \ref{fig:SWCNN}.

\subsection{Document Embeddings}

Doc2Vec \cite{le2014distributed} is an unsupervised approach to generate vector representations for slightly larger bodies of text, such as sentences, paragraphs and documents. It has been adapted from Word2Vec \cite{MikolovEEWRVS} which is used to generate vectors for words in large unlabeled corpora. The vectors generated by this approach come handy in tasks like calculating similarity metrics for sentences, paragraphs and documents. In sequential models like RNNs, the word sequence is captured in the generated sentence vectors. However, in Doc2Vec, the representations are order independent. We use GenSim \cite{rehurek_lrec} to learn 300 dimensional Doc2Vec embeddings for each target description and post title available.

\subsection{Bidirectional LSTM with Attention}

Recurrent Neural Network (RNN) is a class of artificial neural networks which utilizes sequential information and maintains history through its intermediate layers. A standard RNN has an internal state whose output at every time-step which can be expressed in terms of that of previous time-steps. However, it has been seen that standard RNNs suffer from a problem of vanishing gradients \cite{hochreiter1997long}. This means it will not be able to efficiently model dependencies and interactions between sub-word representations that are a few steps apart. LSTMs are able to tackle this issue by their use of gating mechanisms. We convert each article headline into its corresponding sub-word level representation to act as input to our bidirectional LSTMs.

$(\overrightarrow{h}_1, \overrightarrow{h}_2, \dots, \overrightarrow{h}_R)$ represent forward states of the LSTM and its state updates satisfy the following equations:
\begin{equation}
    \big[\overrightarrow{f_t},\overrightarrow{i_t},\overrightarrow{o_t}\big] = \sigma \big[ \overrightarrow{W} \big[\overrightarrow{h}_{t-1},\overrightarrow{r_t}\big] + \overrightarrow{b}\big]
\end{equation}
\begin{equation}
    \overrightarrow{l_t} = \tanh \big[\overrightarrow{V} \big[\overrightarrow{h}_{t-1}, \overrightarrow{r_t}\big] + \overrightarrow{d}\big]
\end{equation}
\begin{equation}
    \overrightarrow{c_t} = \overrightarrow{f_t} \cdot \overrightarrow{c}_{t-1} + \overrightarrow{i_t} \cdot \overrightarrow{l_t}
\end{equation}
\begin{equation}
    \overrightarrow{h_t} = \overrightarrow{o_t} \cdot \tanh(\overrightarrow{c_t})
\end{equation}
here $\sigma$ is the logistic sigmoid function, $\overrightarrow{f_t}$, $\overrightarrow{i_t}$, $\overrightarrow{o_t}$ represent the forget, input and output gates respectively. $\overrightarrow{r_t}$ denotes the input at time $t$ and $\overrightarrow{h_t}$ denotes the latent state, $\overrightarrow{b_t}$ and $\overrightarrow{d_t}$ represent the bias terms. The forget, input and output gates control the flow of information throughout the sequence. $\overrightarrow{W}$ and $\overrightarrow{V}$ are matrices which represent the weights associated with the connections.

 $(\overleftarrow{h}_1, \overleftarrow{h}_2, \dots, \overleftarrow{h}_R)$ denote the backward states and its updates can be computed similarly.
 
\begin{figure}[h]
    \centering
    \includegraphics[height=8cm,width=8cm]{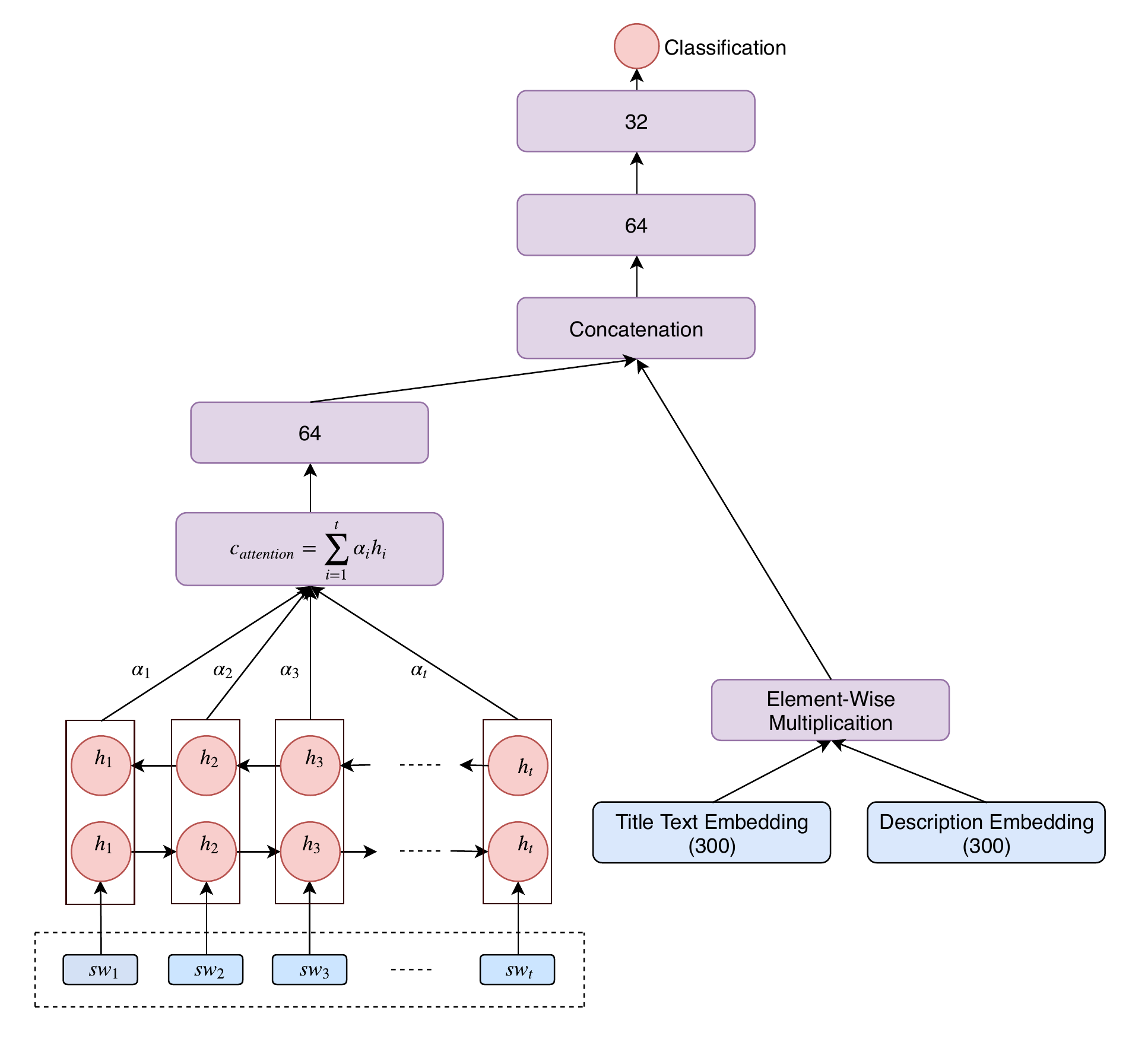}
    \caption{Full Model Architecture}
    \label{fig:full_archi}
\end{figure}

 The number of bidirectional LSTM units is set to a constant \textit{K}, which is the maximum length of all title lengths of records used in training. The forward and backward states are then concatenated to obtain $(h_1, h_2, \dots, h_K)$, where
\begin{equation}
    h_i = \begin{bmatrix}
               \overrightarrow{h}_i \\
               \overleftarrow{h}_i
          \end{bmatrix}
\end{equation}
Finally, we are left with the task of figuring out the significance of each word in the sequence i.e. how much a particular sub-word representation influences the clickbait-y nature of the post. The effectiveness of attention mechanisms have been proven for the task of neural machine translation \cite{bahdanau2014neural} and it has the same effect in this case. The goal of attention mechanisms in such tasks is to derive context vectors which capture relevant source side information and help predict the current target representation. The sequence of annotations generated by the encoder to come up with a context vector capturing how each sub-word contributes to the record's clickbait quotient is of paramount importance to this model. In a typical RNN encoder-decoder framework \cite{bahdanau2014neural}, a context vector is generated at each time-step to predict the target sub-word. However, we only need it for calculation of context vector for a single time-step.
\begin{equation}
c_{attention} = \sum_{j=1}^{K}\alpha_jh_j
\end{equation}
where, $h_1$,\dots,$h_K$ represents the sequence of annotations to which the encoder maps the post title vector and each $\alpha_j$ represents the respective weight corresponding to each annotation $h_j$. This is represented as the left most component in Figure \ref{fig:full_archi}.

\subsection{Doc2Vec Enrichment}

Each record in the dataset has a target description attached with it. This is the entire text of the article whose title has been given. By definition, clickbait articles differ from the content described in their headline. We generate document embeddings for both the title and the article text and perform element wise multiplication over the two. This allows us to capture the interaction between the two, something which has not been used before. Since the title is supposed to mislead the reader with respect to the content, modeling this interaction in terms of their similarity gives an added dimenstion to our approach. It augments the output obtained from the first component.

\subsection{Fusion of Components}

The outputs from the aforementioned components are now concatenated and passed through two dense layers and finally goes into a fully connected layer. This layer finally gives out the probability that a post can be marked clickbait.

\subsection{Learning the Parameters}

We use binary cross-entropy as the loss optimization function for our model. The cross-entropy method \cite{deBoer2005} is an iterative procedure where each iteration can be divided into two stages:

(1) Generate a random data sample (vectors, trajectories etc.) according to a specified mechanism.

(2) Update the parameters of the random mechanism based on the data to produce a "better" sample in the next iteration.

\section{Evaluation Results}

\cite{potthast:2017b} crowdsourced the annotation of 19538 tweets they had curated, into various levels of their clickbait-y nature. These tweets contained the title and text of the article and also included supplementary information such as target description, target keywords and linked images. We trained our model over 17000 records in the described dataset and test it over 2538 disjoint instances from the same. We performed our experiments with the aim of increasing the accuracy and F1 score of the model. Other metrics like mean squared error (MSE) were also considered.

\subsection{Training}

We randomly partition the training set of over 17000 posts into training and validation set in a 4:1 ratio. This ensures that the two sets do not overlap. The model hyperparameters are tuned over the validation set. We initialise the fully connected network weights with the uniform distribution in the range $-\sqrt{{6}/{(fanin + fanout)}}$ and $\sqrt{{6}/{(fanin + fanout)}}$ \cite{glorot2010understanding}. We used a batch size of 256 and adadelta \cite{zeiler2012adadelta} as a gradient based optimizer for learning the model parameters.

\subsection{Model Comparison}

In Table 1, we evaluate our model against the existing state-of-the-art for the dataset used and other models which have employed similar techniques to accomplish the task. It is clear that our proposed model outperforms the previous feature engineering benchmark and other work done in the field both in terms of F1 score and accuracy of detection. Feature engineering models rely on a selection of handcrafted attributes which may not be able to consider all the factors involved in making a post clickbait. The approach proposed in \cite{PhilippeThomasCINN} takes into account each of the textual features available in an individual fashion, considering them to be independent of each other, which is not the case since, by definition of clickbait, the content of the article title and text are not mutually exclusive. \cite{ClickbaitPatent} proposed the integration of multimodal embeddings. \cite{ankesh2017weused} utilise word and character embeddings which do not capture morpheme-level information that may incorporate a surprise element.

\begin{table}
\caption{Model Performance Comparison}
\begin{tabular}{|c|c|c|} \hline
\textbf{Model}&\textbf{F1 Score}&\textbf{Accuracy}\\ \hline
Proposed Approach & \textbf{0.63}& \textbf{83.49}\%\\ \hline
BiLSTM \cite{ankesh2017weused} & 0.61& 83.28\%\\ \hline
Feature Engineering SotA \cite{potthast:2016} & 0.55& 83.24\%\\ \hline
Concatenated NN Architecture \cite{PhilippeThomasCINN} & 0.39& 74\%\\
\hline\end{tabular}
\end{table}

\section{Conclusion}

We have devised an approach to detecting clickbait that puts emphasis on utilising the linguistic value of words by learning its morphological features through its sub-word representations. These embeddings and their dependencies are, in turn, modeled by the LSTM. Attention mechanism allows us to understand the importance of individual representations towards the nature of the post. Using the document embeddings for title and article text allows us to augment the generated embeddings and use as input to a neural network to finally classify the post. In the future, we would like to explore the possibility of integrating the sub-word representations with deep neural networks to better model the temporal and sequential properties of text.

\begin{raggedright}
\bibliography{clickbait17-notebook-lit}
\end{raggedright}
\end{document}